\documentclass[amsmath,amssymb,aps,prb,superscriptaddress,preprint]{revtex4-1}

\usepackage{mathtools}
\usepackage{mathrsfs, amsmath, wasysym}
\usepackage[mathscr]{eucal}
\usepackage{xfrac}
\usepackage{graphicx}
\usepackage{dcolumn}
\usepackage{bm}
\usepackage{color}
\usepackage{tabularx}
\usepackage{natbib}
\usepackage[colorlinks,linkcolor=blue,citecolor=blue,urlcolor=blue]{hyperref}
\usepackage[normalem]{ulem}
\usepackage[all]{hypcap}
\usepackage[dvipsnames,usenames,table]{xcolor}
\usepackage{soul}

\newcommand{\nocontentsline}[3]{}
\newcommand{\tocless}[2]{\bgroup\let\addcontentsline=\nocontentsline#1{#2}\egroup}

\newcommand{\be}{\begin{equation}}
\newcommand{\ee}{\end{equation}}
\newcommand{\beg}{\begin{gather}}
\newcommand{\eeg}{\end{gather}}
\newcommand{\beq}{\begin{eqnarray}}
\newcommand{\eeq}{\end{eqnarray}}
\newcommand{\bea}{\begin{align}}
\newcommand{\eea}{\end{align}}
\newcommand{\beqq}{\begin{eqnarray*}}
\newcommand{\eeqq}{\end{eqnarray*}}

\begin{document}

\title{Sr-induced dipole scatter in BST: Insights from MD simulations using a transferable bond valence-based interatomic potential}

\author{Robert B. Wexler}
\affiliation{Department of Chemistry, University of Pennsylvania, Philadelphia, Pennsylvania 19104, USA}

\author{Yubo Qi}
\affiliation{Department of Physics and Astronomy, Rutgers University, Piscataway, New Jersey 08854, USA}

\author{Andrew M. Rappe}
\affiliation{Department of Chemistry, University of Pennsylvania, Philadelphia, Pennsylvania 19104, USA}

\begin{abstract}
In order to design next-generation ferroelectrics, a microscopic understanding of their macroscopic properties is critical. One means to achieving an atomistic description of ferroelectric and dielectric phenomena is classical molecular dynamics simulations. Previously, we have shown that interatomic potentials based on the bond valence molecular dynamics (BVMD) method can be used to study structural phase transitions, ferroelectric domain nucleation, and domain wall migration in several perovskite oxides and fixed-composition binary and ternary alloys. Most modern devices, however, use variable-composition perovskite oxide alloys such as Ba$_x$Sr$_{1-x}$TiO$_3$ (BST). In this paper, we extend our bond valence approach to BST solid solutions and, in so doing, show that the potential parameters for each element are transferable between materials with different $x$. Using this potential, we perform BVMD simulations investigating the temperature and composition dependence of the lattice constants, Ti displacements, and ferroelectric polarization of BST and find that our predictions match experiments and first-principles theory. Additionally, based on a detailed analysis of local dipole distributions, we demonstrate that substitution of Sr for Ba scrambles dipoles, reduces global polarization, and enhances the order-disorder character of the ferroelectric-paraelectric phase transition.
\end{abstract}

\date{\today}

\maketitle

\section{Introduction \label{sec:intro}}

Bond valence-based interatomic potentials have proven to be a powerful tool, enabling fast and large-scale molecular dynamics (MD) simulations of ferroelectric oxides. Interatomic potentials for several technologically important perovskite materials, such as BaTiO$_3$, PbTiO$_3$, PbZrO$_3$ and BiFeO$_3$ have been successfully developed.~\cite{Cooper03p220,Shin05p054104,Shin08p1206,Liu13p102202,Liu13p104102,Qi16p134308} In addition, bond-valence MD (BVMD) potentials successfully describe PZT Pb(Zr$_{0.5}$Ti$_{0.5}$)O$_{3}$ and 25\% PMN-PT Pb(Mg$_{0.25}$Ti$_{0.25}$Nb$_{0.5}$)O$_{3}$, single-composition binary and ternary alloys.~\cite{Grinberg02p909,Cooper03p220,Shin05p054104,Shin08p1206,Liu13p102202,Liu13p104102,Takenaka13p147602,Qi16p134308,Takenaka17p391,Takenaka18p6567,Kim19p1901060} In these potentials, however, the parameters corresponding to a specific element are material-dependent, not just species-dependent. For example, the fitted Coulombic charges for oxygen are different in each material, which makes BVMD simulations for arbitrary $x$ of $\left( A_{x} A'_{1-x} \right) B {\rm O}_{3}$ or $A \left( B_{x} B'_{1-x} \right) {\rm O}_{3}$ alloys impossible. Given this limitation, we aim to develop transferable interatomic potentials where the parameters corresponding to a particular element depend only on that element's intrinsic properties and, therefore, can be used for different materials and their heterostructures and alloys with diverse chemical compositions. In this paper, we report the successful development of a transferable interatomic potential for Ba$_{x}$Sr$_{1-x}$TiO$_{3}$ (BST). BVMD simulations using this potential accurately reproduce the experimentally determined temperature-composition phase diagram of BST. Additionally, the atomistic nature of our interatomic potential enables analysis of local dipole distributions, which reveal that the substitution of Sr for Ba weakens the correlation between and promotes arbitrarily-oriented Ti displacements, suppresses the global polarization, and changes the character of the ferroelectric-paraelectric phase transition. This work not only presents proof of concept for the development of element-dependent, transferable BVMD interatomic potentials but also provides atomistic insights into the relationship between the thermodynamic properties of BST and its composition.

\section{Computational methods \label{sec:methods}}

\subsection{Bond valence-based interatomic potentials \label{sec:bv}}

\begin{figure}
    \centering
    \includegraphics[width=0.48\textwidth]{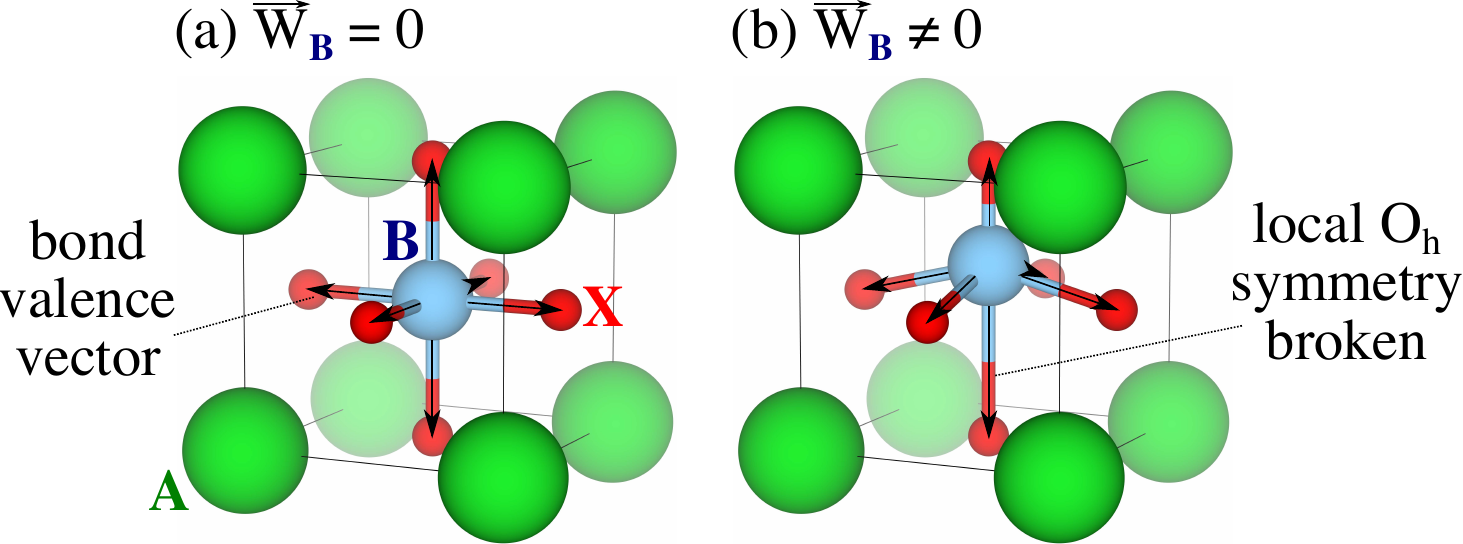}
    \caption{Bond valence vector sum (BVVS) of (a) a centrosymmetric perovskite structure with zero BVVS and (b) a polar perovskite structure with non-zero BVVS. The Ba, Ti, and O atoms are represented by green, blue, and red spheres respectively.}
    \label{fig:bvv}
\end{figure}

\begin{figure}
    \centering
    \includegraphics[width=0.25\textwidth]{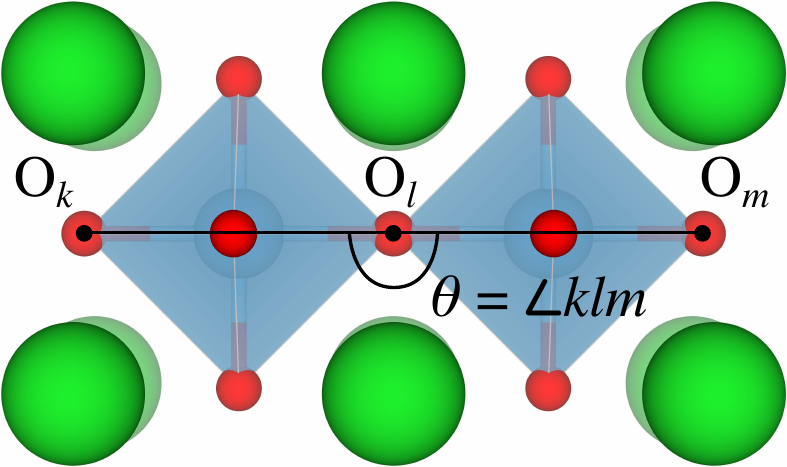}
    \caption{Schematic representation of the angle potential.}
    \label{fig:angle}
\end{figure}

In the bond valence model, the valence of a bond ($V_{ij}$) between atoms $i$ and $j$ is defined as the number of electron pairs used to form that bond and is expressed mathematically as
\begin{equation}
    V_{ij}=\left(\frac{r_{0,ij}}{r_{ij}}\right)^{C_{ij}}
\end{equation}
where $r_{0,ij}$ is a reference distance between atoms $i$ and $j$, $r_{ij}$ is the instantaneous distance, and $C_{ij}$ is an empirical parameter related to the force constant of the chemical bond.~\cite{Brown73p266,Brown76p1957,Grinberg02p909,Shin05p054104,Shin08p015224,Brown09p6858,Liu13p102202,Liu13p104102,Qi16p134308} The valence of an atom $i$ ($V_{i}$) is obtained by summing the bond valences of the bonds it forms with its neighbors $j$,
\begin{equation} \label{eq:atval}
    V_{i}=\sum_j^{N_{n}}V_{ij},
\end{equation}
where $N_{n}$ is the number of neighbors. Due to the bond valence conservation principle, there is an energy penalty ($E_{\rm BV}$) if the atomic valence deviates from its optimal value ($V_{0,i}$)
\begin{equation} \label{eq:bv}
    E_{\rm BV}=\sum_{i}^{N_{a}}S_{i}\left(V_{i}-V_{0,i}\right)^{2}
\end{equation}
where $N_{a}$ is the number of atoms in the unit cell and $S_i$ is a scaling factor. Chemically speaking, $E_{\rm BV}$ describes the energy increase associated with over- and under-coordination of atoms.

The bond valence vector is defined as $\vec{V}_{ij}=V_{ij}\hat{R}_{ij}$ where $\hat{R}_{ij}$ is the unit vector pointing from atom $i$ to atom $j$. The sum of the bond valence vectors of an atom $i$
\begin{equation}
    \vec{W}_{i}=\sum_{j}^{N_{n}}\vec{V}_{ij}
\end{equation}
is a measure of local symmetry breaking. Take, for example, the perovskite $ABX_{3}$ structure shown in Figure~\ref{fig:bvv}. In Figure~\ref{fig:bvv}(a), the $B$-site cation sits at the center of an octahedral cage of $X$ anions. In this arrangement, the $B$-$X$ bond valence vectors cancel, yielding $\vec{W}_{B}=0$. If the $B$-site cation displaces off-center, as in Figure~\ref{fig:bvv}(b), the bond valence vectors along the displacement direction no longer cancel, leading to a non-zero $\vec{W}_{B}$. Similar to Equation~\ref{eq:bv}, a bond valence vector energy ($E_{\rm BVV}$) can be written as
\begin{equation} \label{eq:bvv}
    E_{\rm BVV}=\sum_{i}^{N_{a}}D_{i}\left(\vec{W}^2_{i}-\vec{W}^2_{0,i}\right)^{2}
\end{equation}
where $D_{i}$ is a scaling factor and $| \vec{W}_{0,i} |$ is the preferred bond valence vector length. This energy term is important for capturing the equilibrium off-center displacements of ions ($\vec{W}_{0,i}\neq0$) in ferroelectric materials. Previously, we have shown that $E_{\rm BV}$ and $E_{\rm BVV}$ are equivalent to second- and fourth-moment bond order potentials, respectively.~\cite{Pettifor02p33} This equivalence shows that Equations~\ref{eq:bv} and \ref{eq:bvv} have a quantum mechanical foundation.~\cite{Finnis84p45,Harvey06p1038,Brown09p6858,Liu13p102202,Liu13p104102}

In our scheme, the total energy of the system is given by
\begin{equation}
    E=E_{\rm BV}+E_{\rm BVV}+E_{r}+E_{c}+E_{a}
\end{equation}
\begin{equation}
    E_{r}=\sum_{i<j}\left(\frac{B_{ij}}{r_{ij}}\right)^{12}
\end{equation}
\begin{equation}
    E_{c}=\sum_{i<j}\frac{q_{i}q_{j}}{r_{ij}}
\end{equation}
\begin{equation}
    E_{a}=k\sum_{i}\left(\theta_{i}-180^{\circ{}}\right)^{2}
\end{equation}
where $E_{r}$ is the short-range repulsion energy, $E_{c}$ is the Coulomb energy, $E_{a}$ is the $X_{6}$ octahedral tilting energy, $B_{ij}$ is the short-range repulsion parameter, $q$ is the charge in units of $|e|$, $k$ is a scaling factor, and $\theta_{i}$ is the octahedral tilting angle (see Figure~\ref{fig:angle}) in degrees calculated as $\angle klm$.

Bond valence-based interatomic potentials enable efficient, large-scale MD simulations and have been used successfully in the past to study structural phase transitions,~\cite{Cooper03p220,Shin05p054104,Shin08p1206,Liu13p102202,Liu13p104102,Qi16p134308} the nucleation of ferroelectric domains,~\cite{Shin07p881,Liu16p360,Lu17p222903} the dynamics of the walls separating these domains in perovskites,~\cite{Liu16p360} and relaxor ferroelectrics.~\cite{Takenaka13p147602,Takenaka17p391,Takenaka18p6567,Kim19p1901060} For this reason, we have developed interatomic potentials for several technologically important perovskites such as PbTiO$_{3}$,~\cite{Cooper03p220,Shin05p054104,Shin08p1206,Liu13p104102,Lu17p222903} BiFeO$_{3}$,~\cite{Liu13p102202,Chen15p6371,Agrawal19p034410} PbZrO$_{3}$, BaTiO$_{3}$ (BTO),~\cite{Qi16p134308} PZT,~\cite{Grinberg02p909,Cooper03p220} and PMN-PT.~\cite{Takenaka13p147602,Takenaka17p391,Takenaka18p6567,Kim19p1901060} Based on these successes, we extend this approach to include the perovskite alloy family BST. We note that our previously developed interatomic potential for BTO accurately reproduces many physical properties such as lattice constants, permittivities, and the structural phase transition sequence. Therefore, the parameters related to Ba, Ti, and O elements are kept fixed, and only those related to Sr ($r_{0,{\rm SrO}}$, $C_{\rm SrO}$, $S_{\rm Sr}$, $D_{\rm Sr}$, and $B_{\rm SrX}$ where X $\in$ \{Ba, Sr, Ti, O\}) were optimized. Additionally, in order to maintain charge neutrality, we set the charge of Sr equal to that of Ba.

\subsection{Parameterization \label{sec:dft}}

\begin{table*}
    \centering
     \begin{tabular}{c c c c c c c c c c c c}
        \hline
        \hline
        & & & & & & & $B_{\beta\beta^{\prime}}$ (\AA) & & & \\
        \cline{7-10}
        & $r_{0,\beta{\rm O}}$ & $C_{0,\beta\rm{O}}$ & $q_{\beta}$ ($e$) & $S_{\beta}$ (eV) & $D_{\beta}$ & Ba & Sr & Ti & O & $V_{0,\beta}$ & $\vec{W}_{0,\beta}$ \\
        \hline
        Ba & 2.290 & 8.94 &  1.34730 & 0.59739 & 0.08429 & 2.44805 & 2.40435 & 2.32592 & 1.98792 & 2.0 & 0.11561 \\
        Sr & 2.143 & 8.94 &  1.34730 & 0.63624 & 9.99121 &         & 0.38947 & 1.68014 & 1.96311 & 2.0 & 0.00000 \\
        Ti & 1.798 & 5.20 &  1.28905 & 0.16533 & 0.82484 &         &         & 2.73825 & 1.37741 & 4.0 & 0.39437 \\
        O  &       &      & -0.87878 & 0.93063 & 0.28006 &         &         &         & 1.99269 & 2.0 & 0.31651 \\
        \hline
        \hline
    \end{tabular}
    \caption{Optimized parameters of the bond valence-based interatomic potential for BST. The scaling constant $k$ is 0.0609 eV/$\left( {\rm deg} \right)^2$.}
    \label{tab:params}
\end{table*}

The parameters of the interatomic potential were fit to reproduce density functional theory (DFT)~\cite{Hohenberg64pB864,Kohn65pA1133} calculations of BST using an optimization protocol described elsewhere.~\cite{Liu13p104102} DFT calculations were carried out using the {\sc Quantum ESPRESSO} software package.~\cite{Giannozzi09p395502} Designed,~\cite{Ramer99p12471} optimized, norm-conserving pseudopotentials~\cite{Rappe90p1227} were used to replace the core electrons with a smoother, effective potential. The exchange-correlation contribution to the total energy was calculated using the PBEsol functional, which was designed specifically for bulk solids and provides excellent agreement with the experimental lattice parameters and spontaneous polarization of BTO.~\cite{Perdew08p136406} The electronic wave functions were expanded in a plane-wave basis set with an energy cutoff of 60 Ry. Integrals over the Brillouin zone were evaluated using a $\Gamma$-centered, 4$\times$4$\times$4 $k$-point mesh. Our database consists of 612 structures extracted from variable-cell relaxations; every structure is a 2$\times$2$\times$2 supercell containing 40 atoms. The total energy, force, and pressure convergence criteria for these relaxations were 1.4$\times$10$^{-5}$~eV/supercell, 2.7$\times$10$^{-4}$~eV/\AA, and 0.5 kbar, respectively. For self-consistent field calculations, the total energy convergence threshold was 1.4$\times$10$^{-8}$ eV/supercell. The average absolute difference between the DFT and MD energies is 1.12$\times$10$^{-3}$~eV/atom. The optimized parameters are listed in Table~\ref{tab:params}.

\subsection{Molecular dynamics simulations} \label{sec:md}

MD simulations were performed using an in-house version of the Large-scale Atomic Molecular Massively Parallel Simulator (LAMMPS)~\cite{Plimpton95p1} that was modified to calculate the bond valence and bond valence vector energies. We investigated five different concentrations of Sr in BST: 10\%, 30\%, 50\%, 70\%, and 90\%. For each concentration, Ba was replaced with Sr randomly. Every structure is a 20$\times$20$\times$20 supercell containing 40,000 atoms. We find this to be more than sufficient (only 10$\times$10$\times$10 is necessary) to converge the structural phase transition sequence of Ba$_{0.9}$Sr$_{0.1}$TiO$_{3}$ (see Figure~S1 in the Supplemental Material). Long-range Coulombic interactions were computed using the particle-particle particle-mesh solver with a desired absolute error in the forces of 1$\times$10$^{-4}$~eV/\AA{}. The cutoff distance for short-range interactions was chosen to be 8~\AA{}. Neighbor lists, containing all atom pairs within 10~\AA{}, were updated every step. The time step for MD simulations was 1 fs. We studied temperatures ranging from 10~K to 170~K. For each temperature, the simulation consisted of three steps: (1) $NVT$ equilibration, (2) $NPT$ equilibration, and (3) $NPT$ sampling. The first step runs for 10~ps and generates positions and velocities sampled from the canonical ensemble using the Nos{\'e}-Hoover thermostat~\cite{Nose84p511,Nose84p255,Hoover85p1695} with a temperature damping parameter of 1~ps. The second step relaxes the volume constraint to sample from the isothermal-isobaric ensemble for 40~ps at 1.01325 bar using the Parrinello-Rahman barostat~\cite{Parrinello80p1196,Parrinello81p7182} with a pressure damping parameter of 5~ps. The third step samples the $NPT$-equilibrated structure for 40~ps, from which thermodynamic time averages can be computed.

\section{Results and Discussion \label{sec:r&d}}

\subsection{Thermodynamic properties of BST\label{sec:properties1}}

\begin{figure}
    \centering
    \includegraphics[width=0.35\textwidth]{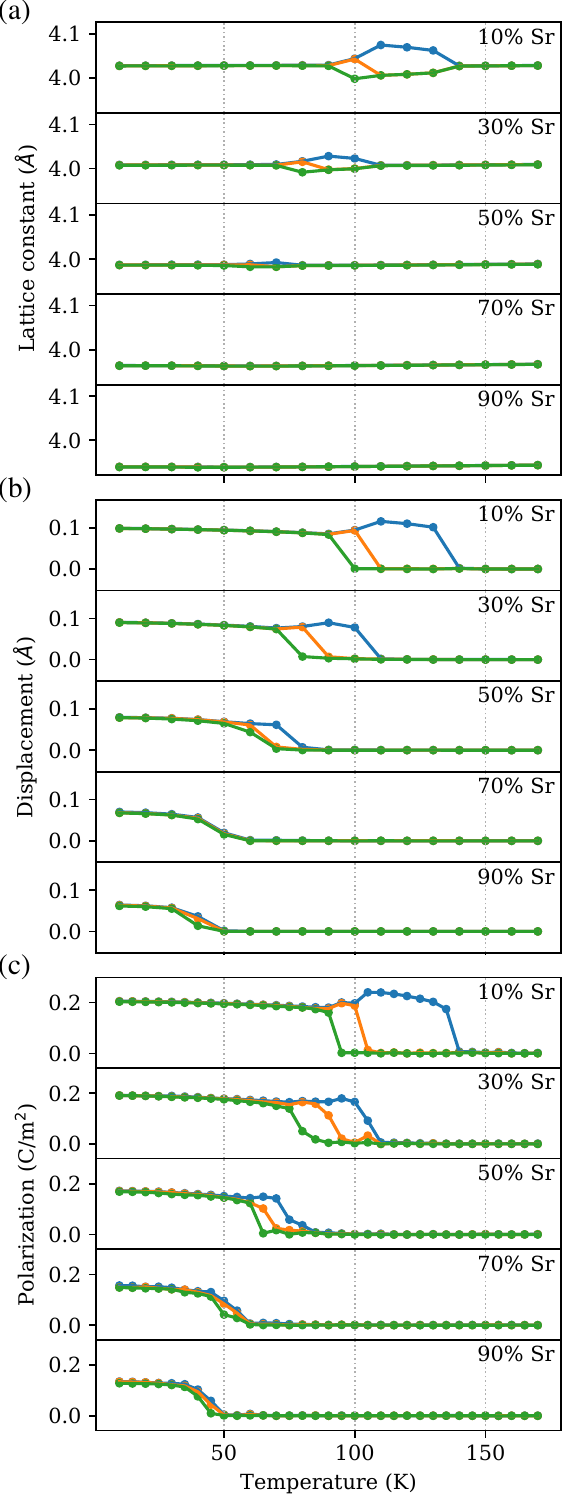}
    \caption{Temperature dependence of the (a) lattice constants, (b) Ti displacements, and (c) polarization components of BST for different compositions.}
    \label{fig:props1-a}
\end{figure}

We tested the performance of the interatomic potential for BST by calculating the lattice constants, components of the Ti displacements, and components of the polarization as a function of the temperature and concentration of Sr. Figure~\ref{fig:props1-a}(a) shows how the lattice constants change as BST is heated from low to high temperature. For 10\% Sr, the evolution of the lattice constants is very similar to that of BTO. It is well-known experimentally that BTO undergoes three structural phase transitions: (1) from rhombohedral ($P \parallel \left[ 111 \right]$) to orthorhombic ($P \parallel \left[ 110 \right]$), (2) from orthorhombic to tetragonal ($P \parallel \left[ 100 \right]$), and (3) from tetragonal to cubic ($P = 0$).~\cite{Lemanov96p3151,Menoret02p224104,Acosta17p041305} First, we will discuss BST with 10\% Sr. At temperatures less than or equal to 90~K, a rhombohedral phase is preferred. Between 90~K and 110~K, BST becomes orthorhombic. From 110~K to 140~K, the system favors a tetragonal crystal structure. Finally, at 140~K, a tetragonal to cubic phase transition occurs. The rhombohedral to orthorhombic to tetragonal to cubic phase transition sequence is in excellent qualitative agreement with experiments and other computational studies.~\cite{Lemanov96p3151,Menoret02p224104,Acosta17p041305,Tinte04p3495,Walizer06p144105,Nishimatsu16p114714} The phase transition temperatures, however, are underestimated, and this is likely due to the fact that DFT underestimates the energy differences between these four structural phases. This hypothesis is supported by the fact that other theoretical models based on DFT also give phase transition temperatures that are too low.~\cite{Tinte99p9679,Tinte04p3495,Walizer06p144105,Nishimatsu16p114714} As the concentration of Sr is increased, the rhombohedral-orthorhombic and tetragonal-cubic phase transition temperatures decrease, and the orthorhombic and tetragonal phases disappear at and above 70\% Sr, both of which are seen experimentally.~\cite{Lemanov96p3151,Menoret02p224104,Acosta17p041305} Additionally, we find that the lattice constants of BST decrease with increasing Sr content because the ionic radius of Sr (1.44~\AA{}) is smaller than that of Ba (1.61~\AA{}).

In addition to the crystal lattice, the position of Ti relative to the center of its O$_{6}$ octahedral cage also depends on the temperature and composition of BST, as shown in Figure~\ref{fig:props1-a}(b). There are four possible displacement modes of Ti: [111], [110], [100], and [000], where the last corresponds to Ti at the center of the octahedron. The transition temperatures for the Ti displacements are the same as those for the structural phase transitions so they will not be repeated here. For 10\% and 30\% Sr, three different displacement transitions are observed: (1) from [111] to [110], (2) from [110] to [100], and (3) from [100] to [000], ranked from the lowest to highest transition temperature. This ordering is the same as that of BTO and is consistent with experimental measurements of the Ti displacement.~\cite{Menoret02p224104} The magnitude of the Ti displacement decreases as more Sr is introduced into the system because the shrinking of the lattice reduces the room for Ti to move off-center.

We also analyzed the effect of temperature and Sr concentration on the polarization of BST. The polarization plotted in Figure~\ref{fig:props1-a}(c) is calculated as
\begin{equation}
    \vec{P} \left( t \right) = \frac{1}{N_{u}} \sum_{i}^{N_{u}} \vec{P}_{i} \left( t \right)
\end{equation}
where $t$ is time, $N_{u}$ is the number of unit cells, and $P_{i}$ is the polarization of unit cell $i$
\begin{equation}
    \vec{P}_{i}\left(t\right)=\frac{1}{\Omega_{i}}\Bigg(\frac{1}{8}\sum_{j=1}^{8}{\rm \bf Z}_{A}^{*}\vec{r}_{A,j}\left(t\right)+{\rm \bf Z}_{\rm Ti}^{*}\vec{r}_{\rm Ti}\left(t\right)+\frac{1}{2}{\rm \bf Z}_{\rm O}^{*}\cdot\sum_{j=1}^{6}\vec{r}_{{\rm O},j}\left(t\right)\Bigg)
\end{equation}
where $\Omega$ is the volume of the unit cell, ${\rm \bf Z}^{*}$ are the Born effective charge tensors (taken from Reference \citenum{Ghosez95p6765}), $A$ is either Ba or Sr, and $\vec{r}$ is the position of each atom relative to the center of the unit cell. Here, we define a unit cell as having eight $A$-site cations at the corners, one Ti at the center, and six O at the face centers. The temperature profile of the polarization components is very similar to that of the Ti displacements. For Sr concentrations less than 30\%, the direction of the polarization changes from [111] to [110] to [100] to [000]. The last change is a ferroelectric to paraelectric phase transition that occurs at the Curie temperature ($T_{C}$). Increasing the amount of Sr reduces the $T_{C}$ substantially, which is in excellent agreement with the experimental literature on BST.~\cite{Lemanov96p3151,Menoret02p224104,Acosta17p041305} The magnitude of the [111] polarization for 10\% Sr (0.35~C/m$^{2}$) also matches previous experimental (0.31~C/m$^{2}$)~\cite{Menoret02p224104} and computational (0.16~C/m$^{2}$ and 0.44~C/m$^{2}$)~\cite{Tinte04p3495,Nishimatsu16p114714} reports.

\subsection{Temperature-composition phase diagram \label{sec:phasediagram}}

\begin{figure}
    \centering
    \includegraphics[width=0.35\textwidth]{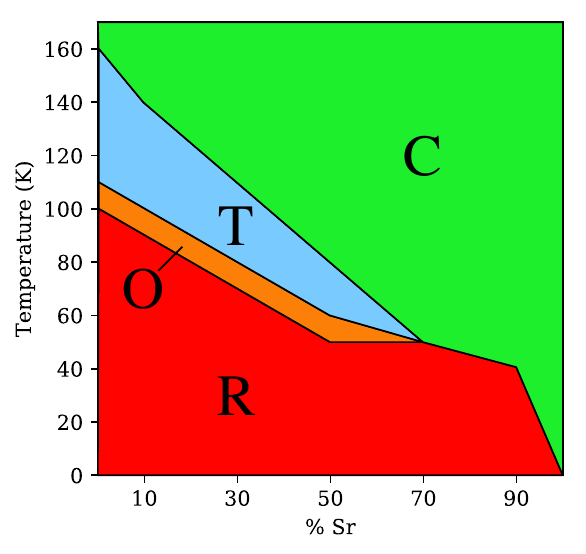}
    \caption{Temperature-composition phase diagram of BST. Red, orange, blue, and green correspond to the rhombohedral, orthorhombic, tetragonal, and cubic phases.}
    \label{fig:phase-diag}
\end{figure}

As a next step in validating our interatomic potential, we construct the temperature-composition phase diagram of BST and compare it with experiments. The phase diagram is shown in Figure~\ref{fig:phase-diag}, where the rhombohedral, orthorhombic, tetragonal, and cubic phases are shaded red, orange, blue, and green, respectively. We are able to reproduce two key features of the phase diagram: (1) the shifts of the ferroelectric-paraelectric and ferroelectric-ferroelectric phase transition lines to lower temperature with increasing Sr concentration and (2) the presence of a tricritical point (TCP) near 70\% Sr. It has been shown that, at the TCP, the character of the ferroelectric-paraelectric phase transition goes from first-order to second-order.~\cite{Lemanov96p3151,Acosta17p041305} This can been seen in Figure~\ref{fig:props1-a}(c) where, between 50\% Sr and 70\% Sr, the transition from non-zero to zero polarization becomes less sharp.

\subsection{Effects of Sr substitution on dipolar structure\label{sec:properties3}}

\begin{figure}
    \centering
    \includegraphics[width=0.7\textwidth]{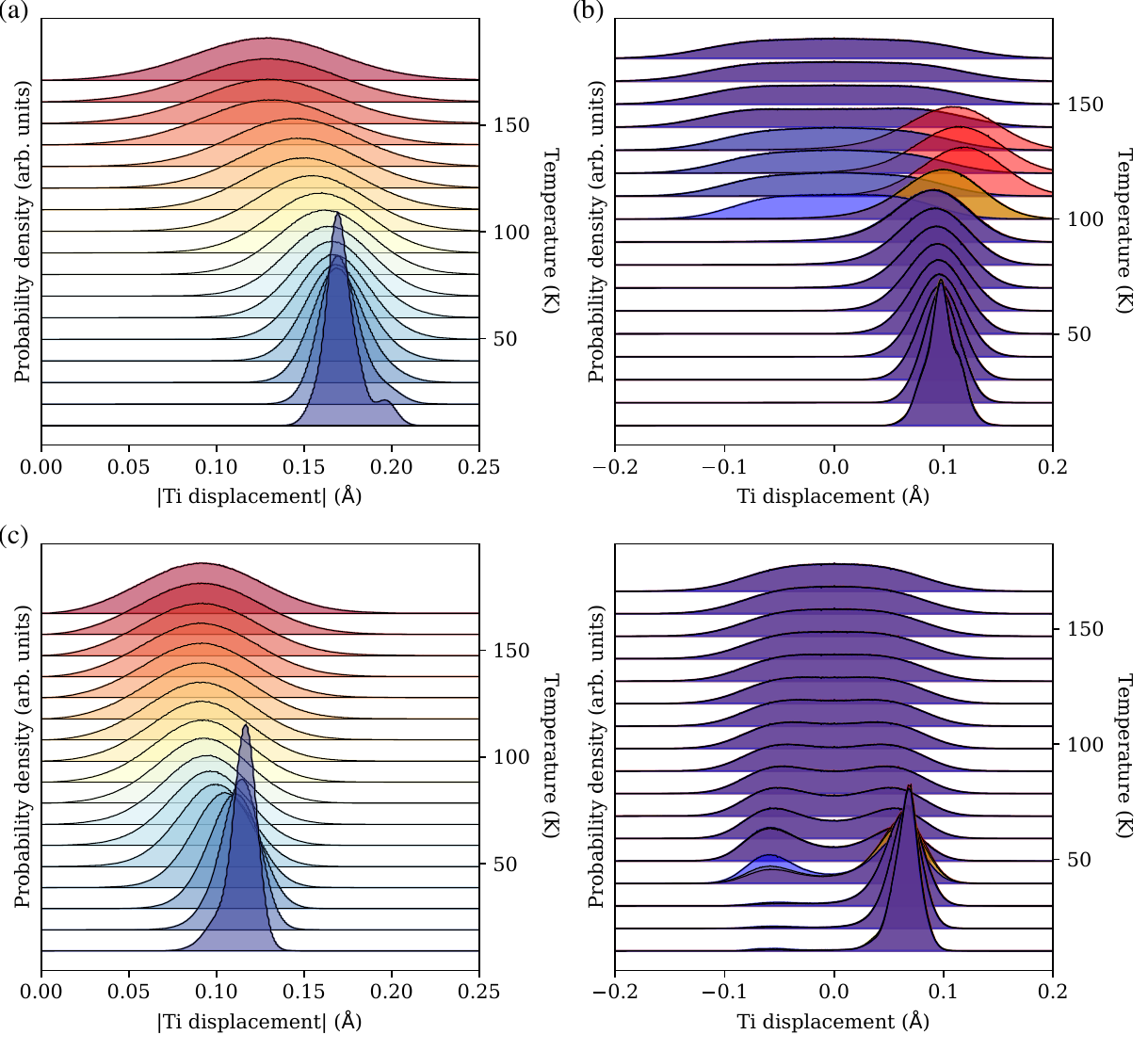}
    \caption{Temperature dependence of (a) $|d_{\rm{Ti}}|$ distributions and (b) Ti displacement distributions in three Cartesian directions for Ba$_{0.9}$Sr$_{0.1}$TiO$_{3}$. Those for Ba$_{0.1}$Sr$_{0.9}$TiO$_{3}$ are shown in (c) and (d), respectively. The colors in (a) and (c) correspond to the temperature of the simulation, blue being the coldest (10 K) and red the hottest (170 K). The colors in (b) and (d) correspond to the components of the Ti displacement in the three Cartesian directions: $x$ is red, $y$ is yellow, and $z$ is blue. If some or all of these distributions overlap, then the resulting color is additive, {\em e.g.} ${\rm red} + {\rm yellow} = {\rm orange}$.}
    \label{fig:props2-a}
\end{figure}

\begin{figure}
    \centering
    \includegraphics[width=0.7\textwidth]{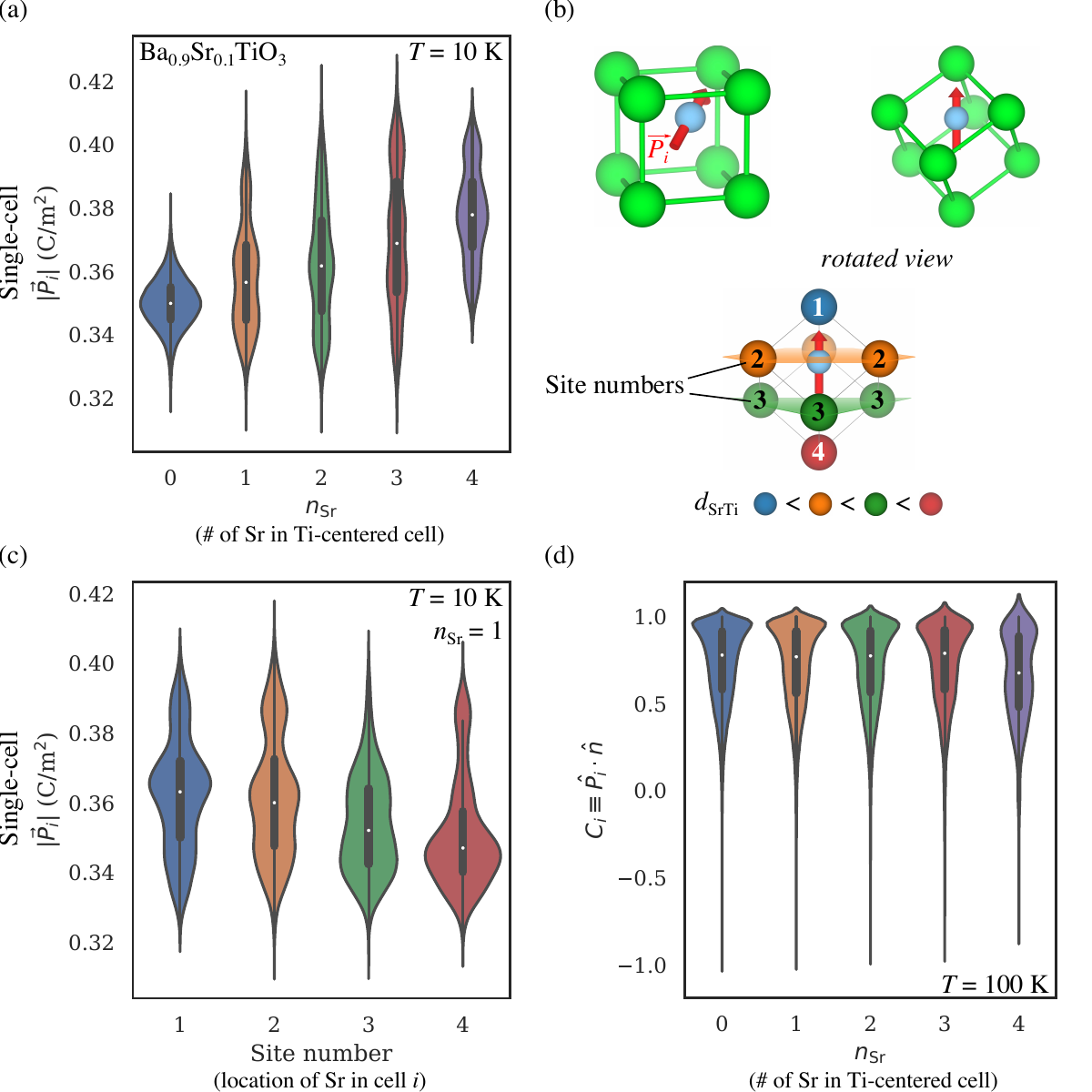}
    \caption{The role of Sr dopants enhancing low-$T$ polarization in Ba$_{0.9}$Sr$_{0.1}$TiO$_{3}$. (a) Dependence of $| P_{i} |$ on $n_{\rm Sr}$ at 10 K. We use a kernel density estimation of the underlying distribution. The bottom and top of the black rectangle correspond to the first and third quartiles, respectively. White points correspond to the median $| P_{i} |$. (b) The top shows the eight nearest-neighbor $A$-sites. The bottom shows the four unique $A$-sites for [111] polarized Ti. (c) Dependence of $| P_{i} |$ on the site number of the $A$-site occupied by Sr for $n_{\rm Sr} = 1$. Numbering corresponds to the bottom of (b). (d) Dependence of the orientational correlation between dipoles, calculated as $C_{i} \equiv \hat{P_{i}} \cdot \hat{n}$, on the number of local Sr at 10 K.}
    \label{fig:pmag10k}
\end{figure}

The strong agreement between BVMD and experimental phase diagrams indicates that this BVMD interatomic potential for BST is accurate and can be used to explore the atomistic origins of structural phase transitions in solid solutions. In order to investigate changes in local physicochemical properties with Sr substitution, we examine the temperature dependence of the Ti displacement distributions of Ba-rich BST (Ba$_{0.9}$Sr$_{0.1}$TiO$_{3}$, see Figures \ref{fig:props2-a}(a) and (b)) and Sr-rich BST (Ba$_{0.1}$Sr$_{0.9}$TiO$_{3}$, see Figures \ref{fig:props2-a}(c) and (d)). Figure~\ref{fig:props2-a}(a) shows that, for Ba-rich BST, the absolute values of the Ti displacements $|d_{\rm{Ti}}|$ are normally distributed at all temperatures except at 10~K where two peaks are observed. The larger peak at 0.17~\AA{} corresponds to the displacement of Ti atoms that are completely surrounded by Ba at the $A$-sites. The smaller peak at 0.20~\AA{}, however, originates from Ti atoms whose neighboring $A$-sites are partially occupied by Sr.~\cite{Tinte04p3495} The effect of Sr on $| d_{Ti} |$ can be seen more clearly in Figure~\ref{fig:pmag10k}(a), where the distribution of $| P_{i} |$ is plotted against the number of Sr in a Ti-centered cell ($n_{\rm Sr}$). Each Ti has eight nearest neighbor $A$-sites and each of these can host either Ba or Sr. Consequently, we define $n_{\rm Sr}$ as the number of Sr occupying nearest neighbor $A$-sites for a particular Ti. We find that as $n_{\rm Sr}$ increases from 0 to   4, the median $| P_{i} |$ (show as white points in Figure~\ref{fig:pmag10k}(a)) increases from 0.350 C/m$^{2}$ to 0.378 C/m$^{2}$. The reason for the increased Ti displacement near Sr is that, for low Sr doping concentrations, the lattice constant does not change, but Sr takes up less volume than Ba. Therefore, the free volume available for off-center Ti displacements is larger when Sr is adjacent. The magnitude of this effect also depends on the position of Sr relative to that of Ti. If Ti is polarized along [111], then there are four unique $A$-sites (see Figure~\ref{fig:pmag10k}(b)). Figure~\ref{fig:pmag10k}(c) shows that, for $n_{\rm Sr} = 1$, as the distance between Sr and Ti increases, the median $| P_{i} |$ decreases from 0.363 C/m$^{2}$ to 0.347 C/m$^{2}$. This suggests that the free volume created by replacing Ba with Sr is more accessible to Ti and subsequently more effective at increasing its polarization when Sr and Ti are closer. Additionally, the multi-peak structure for $n_{\rm Sr} \geq 1$ in Figure~\ref{fig:pmag10k}(a) can be rationalized by the fact that, for different Sr-Ti distances, the $| P_{i} |$ distribution is centered at different values. It can also be seen in Figure~\ref{fig:props2-a}(a) that Ti shifts to smaller displacements with a broader distribution at higher temperatures. The decrease in the mean displacement {\em vs.} temperature is consistent with the Landau-Ginzburg-Devonshire theory of first-order ferroelectric-paraelectric phase transitions.~\cite{Landau37p26,Devonshire49p1040,Ginzburg50p1064} The structural phase transitions of BST also contribute to this decrease by suppressing components of the displacement. On the other hand, the increase in the standard deviation of the Ti displacement magnitude is due to thermal randomization in the presence of disorder. Despite the mean decrease, the distribution remains centered at non-zero displacements even in the paraelectric phase. This suggests that the ferroelectric-paraelectric phase transition of Ba-rich BST has order-disorder character.

The individual phase transitions of Ba-rich BST can be seen more clearly in Figure~\ref{fig:props2-a}(b), which plots the distribution of the Ti displacement components against temperature. The $x$, $y$, and $z$ components are shaded red, yellow, and blue, respectively. For temperatures below 100~K, the three distributions are superimposed, resulting in a single Gaussian (shaded purple). At 100~K, there is a rhombohedral to orthorhombic phase transition that zeroes the mean and flattens the distribution of the $z$ component (shaded blue). The same thing happens to the remaining non-zero components at the higher-temperature phase transitions, $i.e.$ orthorhombic-tetragonal at 110~K and tetragonal-cubic (ferroelectric-paraelectric) at 130~K. This plateauing of the distribution is characteristic of simultaneously order-disorder and displacive phase transitions.~\cite{Qi16p134308} Therefore, the phase transitions of Ba-rich BST exhibit mixed order-disorder and displacive character, which is quite similar to pure BTO.~\cite{Qi16p134308}
 
The distribution of the Ti displacement magnitude for Sr-rich BST is plotted in Figure~\ref{fig:props2-a}(c). At low temperatures (10~K), the peak of the distribution is located at 0.12~\AA{}, which is smaller than those for Ba-rich BST (0.17~\AA{} and 0.20~\AA{}). For Sr-rich BST, the lattice constant is primarily determined by the ionic radius of Sr, which is smaller than that of Ba; this decrease in the lattice constant leads to a reduction in free volume and a suppression of the Ti displacements. The distribution of the Ti displacement components {\em vs.} temperature (see Figure~\ref{fig:props2-a}(d)) show a change in the character of the ferroelectric-paraelectric phase transition that is due to the high concentration of Sr. At low temperatures (10~K), nearly all of the Ti displacements are directed along [111] (shaded purple). There are some antiparallel displacements, however, as indicated by the small peak at -0.06~\AA{}, which suggests that the orientational correlation between dipoles is weakened by the presence of Sr. As the temperature increases (20~K $< T <$ 40~K), the height of the antiparallel peak increases, revealing that the character of the ferroelectric-paraelectric phase transition is predominantly order-disorder. We should also emphasize that, at 40~K, unit cells resembling all three polar phases (rhombohedral (shaded purple), orthorhombic (shaded orange), and tetragonal (shaded blue)) coexist, but due to the lack of strong correlation between dipoles, no single polar phase dominates. At 50~K, the two peaks have equal height, revealing an order-disorder nonpolar state. As the temperature increases further ($T >$ 50~K), the value of the distribution at zero displacement begins to rise, indicating the appearance of displacive character in the phase transition.

In order to analyze the effect of $n_{\rm Sr}$ on the orientation correlation of Ti dipoles, we return to the case of Ba-rich BST, as it allows us to quantify the effect of individual Sr$^{2+}$ cations. Here, we define the orientational correlation of the Ti dipole as
\begin{equation}
    C_{i} \equiv \hat{P_{i}} \cdot \hat{n}
\end{equation}
where $\hat{P_{i}}$ is the local polarization direction, and $\hat{n}$ is the [111] direction. Figure~\ref{fig:pmag10k}(d) shows that as $n_{\rm Sr}$ is increased from 0 to 4, the median $C_{i}$ decreases from 0.726 to 0.659 at 100 K, {\em i.e.} the rhombohedral-orthorhombic phase transition temperature. Consistent with the decrease in the median $C_{i}$, the distribution also shows more dipoles far from [111]. These features suggest that Sr weakens Ti dipole correlations. The extent of $C_{i}$ reduction depends monotonically on $n_{\rm Sr}$. For all $n_{\rm Sr}$, there are a few Ti that are antiparallel to [111], as indicated by the thin but nonzero distributions at negative $C_{i}$. These Ti have undergone thermally-induced, local polarization switching due to the rhombohedral-orthorhombic phase transition. $C_{i}$ analysis enriches our view of the dipolar structure and influence of Sr on Ti by providing quantitative insights about the role of Sr doping in dipole scattering; we observe antiparallel-oriented dipoles for all numbers of local Sr, and a significant weakening of the dipole correlation for $n_{\rm Sr} > 3$.

\section{Conclusions \label{sec:conclusions}}

In conclusion, we have developed a robust interatomic potential for BST based on the bond valence method. This potential enables accurate and efficient large-scale molecular dynamics simulations of ferroelectric alloy phenomena at the atomistic level. Here, we examine the temperature and composition dependence of the lattice parameters, Ti displacements, and polarization, and achieve excellent correspondence with experiment. Additionally, our BST potential is transferable in the sense that the parameters for each element are taken from potentials for other materials that contain the same elements, namely BTO and STO. Such transferability facilitates the construction of potentials for complex perovskite alloy families. Due to the atomistic nature of this potential, we are able to investigate the temperature dependence of the local dipole distributions for both Ba-rich and Sr-rich BST. We discover that the ferroelectric-paraelectric phase transition character of BTO does not change significantly upon 10\% Sr doping. However, in Sr-rich BST, the character of the phase transition is order-disorder at low temperatures, due to the Sr-induced weakening of dipole correlations, with displacive character emerging only at higher temperatures. Looking forward, alloy BVMD potentials not only enable the prediction of macroscopic (lattice constants, polarization, structure phase transitions, and their temperatures) and microscopic (Ti displacements) physical properties of technologically important ferroelectric perovskite alloys such as BST, BZT, PZT, and PMN-PT but also the nanoscale design of new materials via compositional tuning and heterostructure/superlattice engineering.~\cite{Ghosez00p102,Li02p4398,Johnston05p100103,Kaiser99p4657,Tian06p092905,Li07p252904,Bruchhausen08p197402,Kathan10p640,Bruchhausen11p159,Kathan12p052011,Lee13p532,Subramanyam13p13,Vzelezny14p184102,Vzelezny14p929,Wu15p122906,Vzelezny17p214110,Wei05p2140}

\section*{Acknowledgements}
R.B.W. and Y.Q. acknowledge support from the U.S. Office of Naval Research, under Grant No. N00014-17-1-2574. A.M.R. acknowledges support from the DOE Office of Basic Energy Sciences, under Grant No. DE-FG02-07ER46431. The authors also acknowledge computational support from the High-Performance Computing Modernization Office and the National Energy Research Scientific Computing Center.



%


\newpage

\centering \textbf{Supplemental Material}

\setcounter{figure}{0}
\renewcommand{\thefigure}{S\arabic{figure}}
\begin{figure}[h!]
    \centering
    \includegraphics[width=\textwidth]{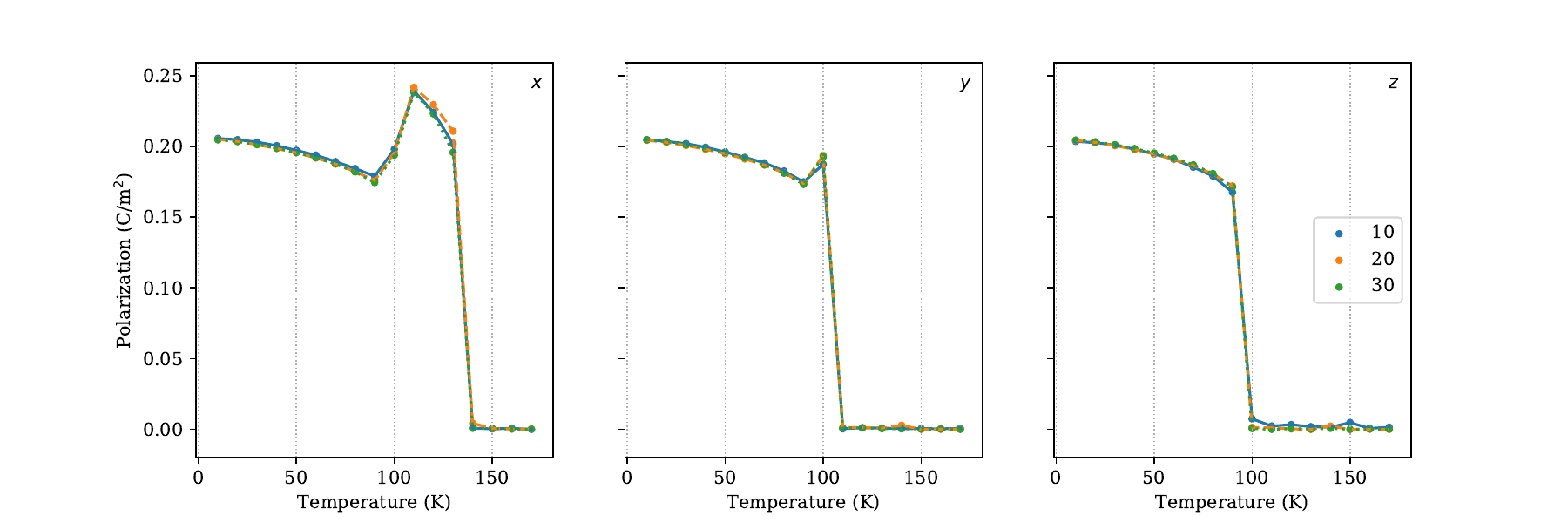}
    \caption{Temperature dependence of the polarization components of BST for different supercell sizes: 10$\times$10$\times$10, 20$\times$20$\times$20, and 30$\times$30$\times$30.}
\end{figure}

\end{document}